\documentclass[a4paper,aps,prd,10pt,preprintnumbers,showpacs,showkeys,twocolumn,superscriptaddress,nofootinbib]{revtex4-1}
\usepackage{cmap,orcidlink,graphicx}
\usepackage[utf8]{inputenc}
\usepackage[T1]{fontenc}

\usepackage{amsmath,amssymb,booktabs,hyperref}
\usepackage{xcolor}
\usepackage{epstopdf}
\def\re#1{Re(#1)}
\def\im#1{Im(#1)}
\def\Order#1{{\cal O}\left(#1\right)}

\begin{document}

\title{Long-lived quasinormal frequencies for regular black hole supported by the Einasto profile in the presence of the magnetic field}
\author{Milena Skvortsova \orcidlink{0000-0003-0812-3171}
}
\email{milenas577@mail.ru}
\affiliation{RUDN University, 6 Miklukho-Maklaya St, Moscow, 117198, Russian Federation}

\begin{abstract}
We investigate quasinormal modes, grey-body factors, and absorption cross-sections of a massive scalar field in regular black-hole spacetimes supported by the Einasto density profile. The analysis is performed for $\tilde n=1/2$, $1$, and $5$, where the scalar mass $\mu$ is treated as an effective parameter induced by an external environment. Quasinormal frequencies are computed with high-order WKB expansions and Pad\'e resummation, and are cross-checked by time-domain evolution. We show that increasing the effective mass and varying the Einasto parameters can strongly suppress the damping rate, leading to long-lived modes and clear quasi-resonant behavior. Grey-body factors obtained from direct WKB transmission and from the QNM-based correspondence agree well in the considered regimes, while their differences remain controlled. Using the transmission coefficients, we derive partial and total absorption cross-sections and demonstrate the expected transition from low-frequency suppression to efficient high-frequency absorption. Our results show that regularity of the core together with environmental parameters leaves a noticeable imprint on both ringdown and scattering observables.
Within this setup, the magnetic field acts as the physical agent that controls the effective mass scale and therefore governs how close the system can approach the quasi-resonant regime.
\end{abstract}


\maketitle

\section{Introduction}

Black holes in realistic astrophysical environments are expected to interact with surrounding matter distributions rather than exist in complete isolation. In particular, galactic dark-matter halos may significantly affect the geometry in the vicinity of compact objects. Among various models used to describe such halos, the Einasto density profile has attracted considerable attention \cite{Einasto:1965czb,Einasto:1989}. Originally introduced in the context of galactic dynamics, the Einasto profile provides an accurate description of dark-matter distributions obtained in numerical simulations of structure formation and is widely used in astrophysical modeling \cite{Springel:2008cc,Navarro:2008kc,Roszkowski:2017nbc,Dutton:2014xda,Catena:2009mf,Navarro:1996gj,Bertone:2005xz,Hernquist:1990be}. Embedding black holes into spacetimes supported by such density profiles therefore provides a useful framework for investigating possible environmental effects on black-hole observables \cite{Konoplya:2021ube,Chakraborty:2024gcr,Pezzella:2024tkf,Konoplya:2022hbl,Zhang:2021bdr,Liu:2024bfj,Feng:2025iao,Daghigh:2022pcr,Zhao:2023tyo,Liu:2024xcd,Mollicone:2024lxy,Tovar:2025apz,Pathrikar:2025sin}.

Another important aspect concerns the internal structure of black holes. Classical solutions of general relativity typically possess spacetime singularities, whose physical interpretation remains problematic. This has motivated the construction of regular black-hole models in which curvature invariants remain finite everywhere. Such geometries can arise either from modified gravity theories or from effective matter sources that regularize the central region of the spacetime \cite{Bardeen:1968,Hayward:2005gi,Spina:2025wxb,Zhang:2024ney,Solodukhin:2025opw,Bueno:2024dgm,Bueno:2025tli,Bueno:2024eig,Konoplya:2020ibi,Bronnikov:2006fu,Bronnikov:2003gx,Casadio:2001jg,Konoplya:2024hfg,Simpson:2018tsi,Ansoldi:2008jw,AyonBeato:1998ub,Bronnikov:2024izh,Bronnikov:2005gm,Kazakov:1993ha,Modesto:2008jz,Bonanno:2000ep,Lan:2023cvz,Konoplya:2023ppx,Bonanno:2023rzk,Konoplya:2024kih}. Regular black holes surrounded by astrophysical matter distributions represent an interesting class of solutions that combine two physically motivated ingredients: the removal of the central singularity and the presence of an external environment \cite{Dymnikova:1992ux,Konoplya:2025ect}.

Perturbations of black-hole spacetimes play a central role in understanding their dynamical and observational properties. After a perturbation, a black hole responds through damped oscillations known as quasinormal modes (QNMs) \cite{Kokkotas:1999bd, Konoplya:2011qq, Berti:2009kk, Bolokhov:2025uxz}, which are determined by the parameters of the background geometry and the properties of the perturbing field. These modes are of particular interest in the context of gravitational-wave observations, since the ringdown phase of a compact-object merger is dominated by QNMs. Environmental effects, such as surrounding matter distributions, could therefore produce measurable modifications of the quasinormal spectrum.

    In many astrophysical scenarios black holes are also embedded in strong electromagnetic environments. Magnetic fields may arise from accretion disks, magnetized plasma, or large-scale galactic structures (see \cite{Rayimbaev:2021vsq,Shaymatov:2022enf,Al-Badawi:2025kbi}  and references therein). The presence of an external magnetic field can influence the dynamics of fields propagating near the black hole. In particular, a massless field may acquire an effective mass term due to its interaction with the magnetic background \cite{Wu:2015fwa, Konoplya:2008hj,Huang:2015cha,Lopez:2022uie} or tidal force from extra dimensions \cite{Seahra:2004fg}. 
The varying the ambient magnetic-field strength provides a direct astrophysical knob that can mimic or amplify environmental effects usually attributed only to matter distributions. In our parameterization this knob is encoded by $\mu$, so the trends we report for increasing $\mu$ can be read as trends expected for stronger magnetic environments (for fixed azimuthal number $m$).
Such effective masses have been discussed in various contexts and provide an additional parameter that can modify the propagation of perturbations and the resulting quasinormal spectrum \cite{Burikham:2017gdm,  Zhang:2018jgj, Aragon:2020teq, Konoplya:2017tvu, Bolokhov:2023ruj, Zhidenko:2006rs, Gonzalez:2022upu, Ohashi:2004wr, Konoplya:2018qov, Ponglertsakul:2020ufm, Konoplya:2004wg,Skvortsova:2024eqi}. In addition, a massive term drastically changes the character of asymptotic tails transforming them into oscillatory ones with a slowly decaying power-law envelope \cite{Koyama:2001ee, Koyama:2000hj, Gibbons:2008gg, Jing:2004zb, Rogatko:2007zz, Moderski:2001tk, Lutfuoglu:2026fpx}. An important feature of massive modes is related to the existence of long-lived and even arbitrarily long-lived modes, called quasi-resonances, \cite{Ohashi:2004wr, Konoplya:2004wg, Zinhailo:2018ska, Fernandes:2021qvr, Konoplya:2017tvu, Percival:2020skc,Zhidenko:2006rs, Zinhailo:2024jzt}. Finally, massive or effectively massive perturbations could contribute at very large wavelengths in the Pulsar Timing Array experiments \cite{Konoplya:2023fmh, NANOGrav:2023hvm}.

Besides quasinormal modes, another important characteristic of wave propagation in black-hole spacetimes is given by the grey-body factors (GBFs) \cite{Page:1976df,Page:1976ki,Kanti:2014vsa}. These quantities describe the probability for waves generated near the horizon to propagate through the effective potential barrier surrounding the black hole and reach asymptotic observers. Grey-body factors determine the spectrum of Hawking radiation and are therefore essential for understanding the radiative properties of black holes. They are also closely related to the scattering problem for fields in curved spacetime and provide complementary information to that obtained from quasinormal modes.

Perturbations, propagation and spectra of massless fields in the background of the regular black holes sourced by the Einasto profile were considered in \cite{Konoplya:2025ect,Lutfuoglu:2026zel,Bolokhov:2026eqf}, while spectra of regular black holes supported by Dehnen profiles were studied in \cite{Bolokhov:2025fto,Lutfuoglu:2025mqa,Saka:2025xxl}.
In the present work we investigate the quasinormal modes and grey-body factors of a massive scalar field propagating in the background of a regular black hole supported by an Einasto density profile. The scalar field is assumed to be initially massless, while the effective mass term arises due to the presence of an external magnetic field. We analyze how both the environmental matter distribution and the effective mass influence the quasinormal spectrum and the transmission probabilities associated with grey-body factors. In particular, we determine the dependence of the oscillation frequencies and damping rates on the parameters of the Einasto profile and on the effective scalar mass, and we study the corresponding modifications of the scattering properties of the field.

The paper is organized as follows. In Sec.~\ref{sec:EinastoBackground} we present the regular black-hole background generated by the Einasto density profile. Section~\ref{sec:waveeq} derives the radial equation for massive scalar perturbations and states the quasinormal-mode boundary conditions. In Sec.~\ref{sec:methods} we briefly summarize the numerical tools used for quasinormal frequencies and transmission coefficients. The results for the quasinormal spectra and grey-body factors are then reported in Secs.~\ref{sec:QNMs} and \ref{sec:gbf}, respectively, including the corresponding absorption cross-sections. Finally, Sec.~\ref{sec:conc} contains our concluding remarks.

\section{Regular black holes supported by the Einasto profile}
\label{sec:EinastoBackground}

We consider static spherically symmetric configurations described by
\begin{equation}
ds^2=-f(r)\,dt^2+\frac{dr^2}{f(r)}+r^2 d\sigma^2 ,
\end{equation}
where the metric function is written in terms of the mass distribution $m(r)$,
\begin{equation}
f(r)=1-\frac{2M(r)}{r}.
\end{equation}
The spacetime is sourced by an effective anisotropic fluid with energy density
$\rho(r)$ and radial pressure $P_r(r)$. Following the construction of
Ref.~\cite{Konoplya:2025ect}, we impose
\begin{equation}
P_r=-\rho ,
\end{equation}
which simplifies the Einstein equations and yields
\begin{equation}
M(r)=4\pi\int_0^r x^2\rho(x)\,dx .
\end{equation}
If the density remains finite at the origin, the geometry is regular and all curvature invariants remain bounded.

As the matter source we adopt the Einasto distribution
\begin{equation}
\rho(r)=\rho_0\exp\!\left[-\left(\frac{r}{h}\right)^{1/\tilde n}\right],
\qquad \tilde n>0 ,
\end{equation}
commonly used to describe dark-matter halos. The parameter $h$ sets the characteristic radial scale, while the index $\tilde n$ controls the slope of the density falloff. The total mass
\begin{equation}
M=\lim_{r\to\infty}M(r)
\end{equation}
is finite for all $\tilde n>0$, so that the spacetime approaches the Schwarzschild solution at large distances.

Near the origin the density tends to a constant value and the metric behaves as
\begin{equation}
f(r)=1-\frac{8\pi\rho_0}{3}\,r^2+\mathcal{O}(r^3),
\end{equation}
which corresponds to a de Sitter–type core. Consequently the central singularity is removed. Depending on the compactness of the distribution (controlled by $h/M$), the spacetime may contain two horizons, a single extremal horizon, or no horizons.

For generic $\tilde n$ the mass function must be evaluated numerically. However, several values allow analytic expressions illustrating the properties of the solution.

For $\tilde n=\tfrac12$ (Gaussian density) and $\tilde n=1$ (exponential density) the metric functions can be written as
\begin{equation}
f_{\tilde n}(r)=1-\frac{2M}{r}+F_{\tilde n}(r,h,M),
\end{equation}
where
\begin{align}
F_{1/2}(r)&=-\frac{2M}{r}\!\left[\mathrm{erf}\!\left(\frac{r}{h}\right)-1\right]
+\frac{4M}{\sqrt{\pi}h}e^{-r^2/h^2}, \\
F_{1}(r)&=M\,\frac{2h^2+2hr+r^2}{h^2 r}e^{-r/h}.
\end{align}

In both cases the central expansion yields
\begin{equation}
f(r)=1-\Lambda_{\rm eff} r^2+\mathcal{O}(r^3),
\end{equation}
with an effective de Sitter curvature scale determined by $\rho_0$. Thus the geometry is regular at the origin. 

The parameter $\tilde n$ primarily controls how extended the matter distribution is. Small values of $\tilde n$ produce rapidly decaying profiles that approach the Schwarzschild metric outside a narrow central region, whereas larger $\tilde n$ correspond to more extended halos and therefore stronger deviations near the horizon.

It is worth noticing that some other distributions of matter (see, for instance, \cite{Gondolo:1999ef,Cardoso:2021wlq}) cannot lead to regular black hole solutions.

In the following sections we investigate wave propagation in these geometries and determine the corresponding quasinormal spectra and grey-body factors.

\begin{figure*}
\centering
\includegraphics[width=0.32\textwidth]{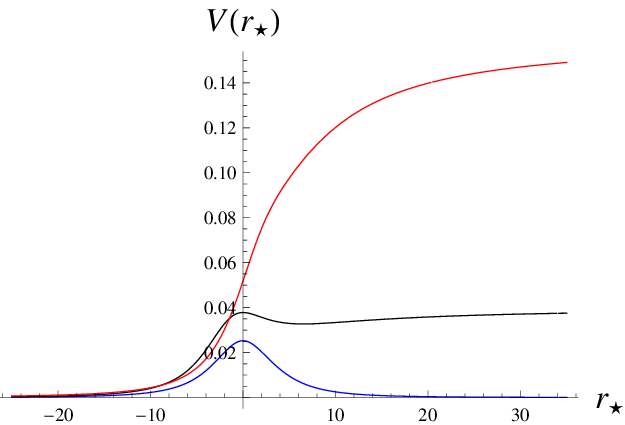}
\hfill
\includegraphics[width=0.32\textwidth]{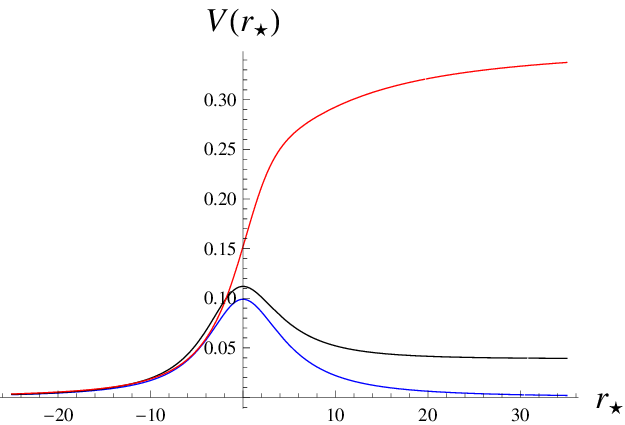}
\hfill
\includegraphics[width=0.32\textwidth]{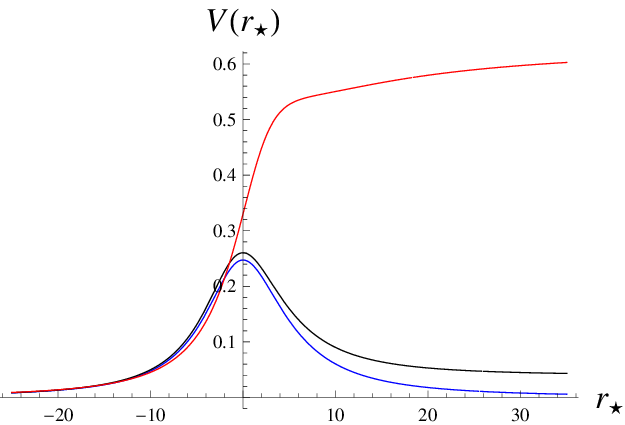}
\caption{Model with $\tilde{n}=1/2$, $h=1.05$. Effective potential as a function of the tortoise coordinate $r^{*}$ for scalar perturbations with $M=1$: left panel $\ell=0$, $\mu=0,0.2,0.4$ (blue, black, red); middle panel $\ell=1$, $\mu=0,0.2,0.6$ (blue, black, red); right panel $\ell=2$, $\mu=0,0.2,0.8$ (blue, black, red).}
\label{fig:potentials_all_nhalf}
\end{figure*}

\begin{figure*}
\centering
\includegraphics[width=0.32\textwidth]{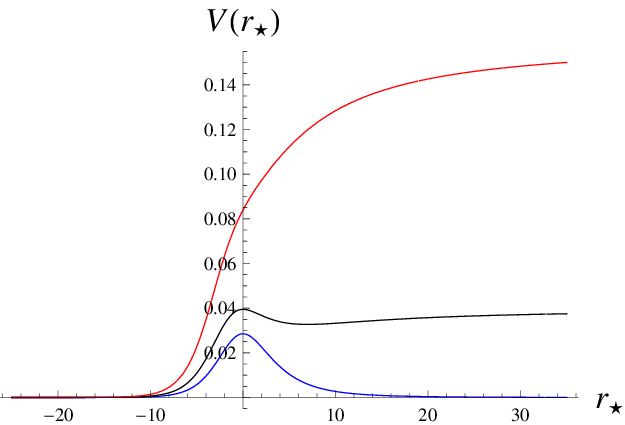}
\hfill
\includegraphics[width=0.32\textwidth]{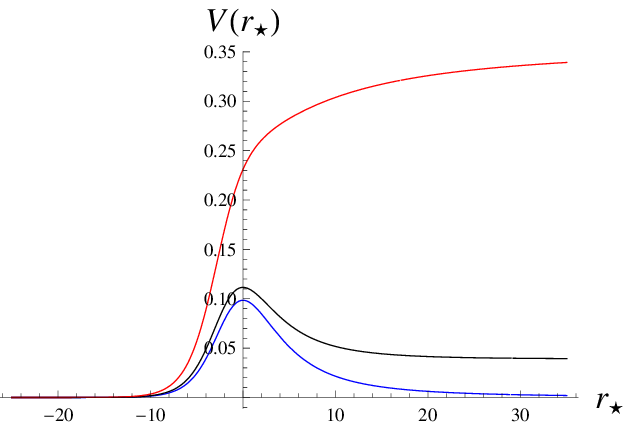}
\hfill
\includegraphics[width=0.32\textwidth]{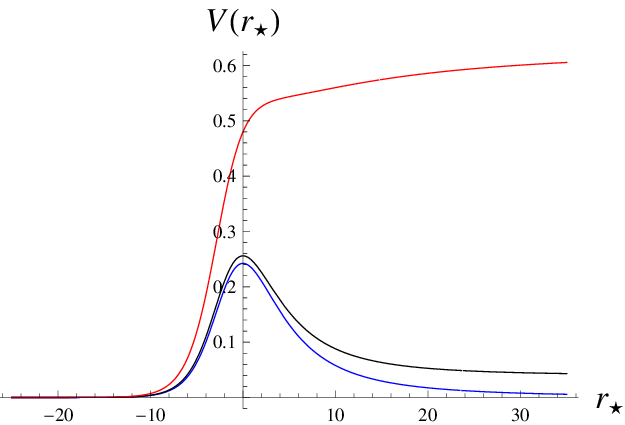}
\caption{Model with $\tilde{n}=1$, $h=0.38$. Effective potential as a function of the tortoise coordinate $r^{*}$ for scalar perturbations with $M=1$: left panel $\ell=0$, $\mu=0,0.2,0.4$ (blue, black, red); middle panel $\ell=1$, $\mu=0,0.2,0.6$ (blue, black, red); right panel $\ell=2$, $\mu=0,0.2,0.8$ (blue, black, red).}
\label{fig:potentials_all}
\end{figure*}

\section{Massive scalar perturbations and boundary conditions}
\label{sec:waveeq}

We study the propagation of a massive test scalar field in the background of a static, spherically symmetric and asymptotically flat black hole. The field satisfies the Klein--Gordon equation
\begin{equation}
(\Box-\mu^2)\Phi=0 ,
\end{equation}
where $\mu$ denotes the scalar mass. Employing the standard decomposition
\begin{equation}
\Phi(t,r,\theta,\phi)=\frac{\Psi(r)}{r}Y_{\ell m}(\theta,\phi)e^{-i\omega t},
\end{equation}
with $Y_{\ell m}$ being spherical harmonics, the equation reduces to a radial wave equation. Introducing the tortoise coordinate
\begin{equation}
\frac{dr_*}{dr}=f^{-1}(r),
\end{equation}
one obtains the Schrödinger-type equation (see  \cite{Bagrov:1990hv, Konoplya:2018arm} for a general approach)   
\begin{equation}
\frac{d^2\Psi}{dr_*^2}+\left(\omega^2-V(r)\right)\Psi=0 .
\end{equation}

For a massive scalar field the effective potential takes the form
\begin{equation}
V(r)=f(r)\left(\frac{\ell(\ell+1)}{r^2}+\frac{f'(r)}{r}+\mu^2\right).
\end{equation}
At large distances the metric approaches the Schwarzschild form and the potential tends to $\mu^2$, so that the asymptotic wave number becomes
\begin{equation}
\chi=\sqrt{\omega^2-\mu^2}.
\end{equation}
Thus the propagation of massive fields differs from the massless case by the presence of a threshold frequency $\omega=\mu$ separating propagating and exponentially decaying modes. We assume $\omega>\mu$, which ensures that $\chi$ is real, and fix its sign to match that of $\omega$.
The scattering problem describes the transmission of waves generated near the horizon through the effective potential barrier. The boundary conditions correspond to a purely ingoing wave at the event horizon,
\begin{equation}
\Psi \to T\,e^{-i\omega r_*}, 
\qquad r_* \to -\infty ,
\end{equation}
while at spatial infinity the solution contains incoming and reflected components,
\begin{equation}
\Psi \to e^{-i\chi r_*}+R\,e^{i\chi r_*},
\qquad r_* \to +\infty .
\end{equation}
Here $R$ and $T$ denote the reflection and transmission amplitudes. Conservation of flux implies
\begin{equation}
|R|^2+\frac{\omega}{\chi}|T|^2=1 .
\end{equation}
The grey-body factor is defined as the transmission probability
\begin{equation}
\Gamma_\ell(\omega)=1-|R|^2 = \frac{\omega}{\chi}|T|^2,
\end{equation}
which measures the fraction of radiation with frequency $\omega$ and multipole number $\ell$ that reaches infinity after being scattered by the potential barrier.

\begin{table}[t]
\centering
\small
\setlength{\tabcolsep}{7pt}
\resizebox{\columnwidth}{!}{%
\begin{tabular}{c c c c}
\toprule
$\mu$ & WKB16 ($\tilde m=8$) & WKB14 ($\tilde m=7$) & Diff. (\%) \\
\midrule
\multicolumn{4}{c}{$\ell=1$} \\
\midrule
$0$ & $0.288295-0.089840 i$ & $0.288057-0.089118 i$ & $0.252$ \\
$0.05$ & $0.289429-0.089189 i$ & $0.289270-0.088580 i$ & $0.208$ \\
$0.10$ & $0.292975-0.087382 i$ & $0.292912-0.086952 i$ & $0.142$ \\
$0.15$ & $0.298992-0.084461 i$ & $0.298986-0.084160 i$ & $0.0970$ \\
$0.20$ & $0.307463-0.080345 i$ & $0.307493-0.080011 i$ & $0.106$ \\
$0.25$ & $0.318406-0.074900 i$ & $0.318456-0.074313 i$ & $0.180$ \\
$0.30$ & $0.332108-0.067580 i$ & $0.332045-0.067554 i$ & $0.0202$ \\
$0.35$ & $0.347647-0.057503 i$ & $0.347641-0.057519 i$ & $0.00486$ \\
$0.45$ & $0.399637-0.031394 i$ & $0.399651-0.031363 i$ & $0.00855$ \\
\midrule
\multicolumn{4}{c}{$\ell=2$} \\
\midrule
$0$ & $0.480184-0.090306 i$ & $0.480163-0.090304 i$ & $0.00424$ \\
$0.05$ & $0.480990-0.090094 i$ & $0.480969-0.090092 i$ & $0.00430$ \\
$0.10$ & $0.483411-0.089455 i$ & $0.483389-0.089452 i$ & $0.00462$ \\
$0.15$ & $0.487456-0.088375 i$ & $0.487429-0.088368 i$ & $0.00573$ \\
$0.20$ & $0.493139-0.086834 i$ & $0.493095-0.086814 i$ & $0.00972$ \\
$0.25$ & $0.500479-0.084801 i$ & $0.500481-0.084624 i$ & $0.0348$ \\
$0.30$ & $0.509504-0.082235 i$ & $0.509572-0.082231 i$ & $0.0132$ \\
$0.35$ & $0.520244-0.079085 i$ & $0.520345-0.079068 i$ & $0.0194$ \\
$0.40$ & $0.532745-0.075287 i$ & $0.532730-0.075277 i$ & $0.00336$ \\
$0.45$ & $0.547044-0.070759 i$ & $0.546988-0.070637 i$ & $0.0245$ \\
$0.50$ & $0.563194-0.065357 i$ & $0.563151-0.065352 i$ & $0.00778$ \\
$0.55$ & $0.581215-0.059052 i$ & $0.581215-0.059052 i$ & $0.00003$ \\
$0.60$ & $0.601062-0.051490 i$ & $0.601158-0.051411 i$ & $0.0204$ \\
$0.65$ & $0.623213-0.042973 i$ & $0.622762-0.043082 i$ & $0.0744$ \\
$0.70$ & $0.643260-0.035168 i$ & $0.646142-0.036008 i$ & $0.466$ \\
\bottomrule
\end{tabular}%
}
\caption{Comparison of the fundamental scalar quasinormal frequency ($n=0$) for the regular black-hole model with $\tilde n=1/2$, $h=1.05$ and $M=1$, computed with 16th- and 14th-order WKB methods using Pad\'e approximants ($\tilde m=8$ and $\tilde m=7$, respectively). Rows are grouped by multipole number $\ell$. The last column reports the relative deviation between the two WKB estimates in percent.}
\label{tab:qnm_wkb_comparison_nhalf}
\end{table}

\begin{table}[t]
\centering
\small
\setlength{\tabcolsep}{7pt}
\resizebox{\columnwidth}{!}{%
\begin{tabular}{c c c c}
\toprule
$\mu$ & WKB16 ($\tilde m=8$) & WKB14 ($\tilde m=7$) & Diff. (\%) \\
\midrule
\multicolumn{4}{c}{$\ell=1$} \\
\midrule
$0$ & $0.286871-0.105160 i$ & $0.286872-0.105160 i$ & $0.0001$ \\
$0.05$ & $0.288037-0.104373 i$ & $0.288037-0.104373 i$ & $0.0000$ \\
$0.10$ & $0.291538-0.102002 i$ & $0.291538-0.102001 i$ & $0.0002$ \\
$0.15$ & $0.297388-0.097986 i$ & $0.297393-0.098003 i$ & $0.0055$ \\
$0.20$ & $0.305652-0.092313 i$ & $0.305660-0.092308 i$ & $0.0028$ \\
$0.25$ & $0.316290-0.084843 i$ & $0.316292-0.084859 i$ & $0.0048$ \\
$0.30$ & $0.329338-0.075457 i$ & $0.329320-0.075383 i$ & $0.0224$ \\
$0.35$ & $0.344642-0.064035 i$ & $0.344465-0.063895 i$ & $0.0645$ \\
$0.40$ & $0.364744-0.049116 i$ & $0.364850-0.048088 i$ & $0.281$ \\
\midrule
\multicolumn{4}{c}{$\ell=2$} \\
\midrule
$0$ & $0.475993-0.104398 i$ & $0.475993-0.104398 i$ & $0.00006$ \\
$0.05$ & $0.476828-0.104078 i$ & $0.476827-0.104078 i$ & $0.00005$ \\
$0.10$ & $0.479332-0.103117 i$ & $0.479332-0.103117 i$ & $0.00005$ \\
$0.15$ & $0.483512-0.101512 i$ & $0.483512-0.101512 i$ & $0.00004$ \\
$0.20$ & $0.489379-0.099257 i$ & $0.489379-0.099256 i$ & $0.00002$ \\
$0.25$ & $0.496946-0.096341 i$ & $0.496946-0.096341 i$ & $0$ \\
$0.30$ & $0.506233-0.092750 i$ & $0.506233-0.092750 i$ & $0.00004$ \\
$0.35$ & $0.517263-0.088464 i$ & $0.517263-0.088464 i$ & $0.00009$ \\
$0.40$ & $0.530066-0.083453 i$ & $0.530064-0.083452 i$ & $0.00032$ \\
$0.45$ & $0.544673-0.077672 i$ & $0.544674-0.077672 i$ & $0.00014$ \\
$0.50$ & $0.561123-0.071059 i$ & $0.561123-0.071059 i$ & $0.00002$ \\
$0.55$ & $0.579450-0.063525 i$ & $0.579451-0.063525 i$ & $0.00007$ \\
$0.60$ & $0.599668-0.054917 i$ & $0.599690-0.054928 i$ & $0.00400$ \\
$0.65$ & $0.621869-0.045143 i$ & $0.621748-0.044916 i$ & $0.0413$ \\
$0.70$ & $0.645212-0.034838 i$ & $0.645236-0.034858 i$ & $0.00477$ \\
\bottomrule
\end{tabular}%
}
\caption{Comparison of the fundamental scalar quasinormal frequency ($n=0$) for a regular Einasto black hole with $\tilde n=1$, $h=0.38$  and $M=1$, computed with 16th- and 14th-order WKB methods using Pad\'e approximants ($\tilde m=8$ and $\tilde m=7$, respectively). Rows are grouped by multipole number $\ell$. The last column reports the relative deviation between the two WKB estimates in percent.}
\label{tab:qnm_wkb_comparison}
\end{table}

\begin{table*}[t]
\centering
\small
\setlength{\tabcolsep}{7pt}
\begin{tabular}{c c c c}
\toprule
$\mu$ & $\ell=0$ & $\ell=1$ & $\ell=2$ \\
\midrule
$0.00$ & $0.151987-0.0942521 i$ & $0.426443-0.0917513 i$ & $0.706753-0.0914151 i$ \\
$0.05$ & $0.152726-0.0931522 i$ & $0.427004-0.0915248 i$ & $0.707116-0.0913288 i$ \\
$0.10$ & $0.154717-0.0899293 i$ & $0.428687-0.0908402 i$ & $0.708204-0.0910690 i$ \\
$0.15$ & $0.157619-0.0848362 i$ & $0.431501-0.0896819 i$ & $0.710022-0.0906334 i$ \\
$0.20$ & $0.160277-0.0794407 i$ & $0.435456-0.0880230 i$ & $0.712574-0.0900180 i$ \\
$0.25$ & $0.187492-0.0658281 i$ & $0.440567-0.0858221 i$ & $0.715867-0.0892169 i$ \\
$0.30$ & -- & $0.446847-0.0830209 i$ & $0.719910-0.0882225 i$ \\
$0.35$ & -- & $0.454307-0.0795400 i$ & $0.724715-0.0870244 i$ \\
$0.40$ & -- & $0.462939-0.0752766 i$ & $0.730295-0.0856097 i$ \\
$0.45$ & -- & $0.472702-0.0701046 i$ & $0.736666-0.0839621 i$ \\
$0.50$ & -- & $0.483445-0.0639432 i$ & $0.743847-0.0820612 i$ \\
$0.55$ & -- & $0.495527-0.0575449 i$ & $0.751860-0.0798812 i$ \\
$0.60$ & -- & $0.518048-0.0393810 i$ & $0.760729-0.0773895 i$ \\
$0.65$ & -- & -- & $0.770480-0.0745450 i$ \\
$0.70$ & -- & -- & $0.781138-0.0712951 i$ \\
$0.75$ & -- & -- & $0.792726-0.0675731 i$ \\
$0.80$ & -- & -- & $0.805259-0.0632957 i$ \\
$0.85$ & -- & -- & $0.818721-0.0583767 i$ \\
$0.90$ & -- & -- & $0.833126-0.0528729 i$ \\
$0.95$ & -- & -- & $0.850393-0.0465785 i$ \\
$1.00$ & -- & -- & $0.877380-0.00869229 i$ \\
\bottomrule
\end{tabular}
\caption{Fundamental scalar quasinormal frequencies ($n=0$) for the Einasto model with $\tilde n=5$, $h=1.5 \times 10^{-6}$ and $M=1$, obtained with the 8th-order WKB method and Pad\'e approximant $\tilde m=4$. Dashes indicate values not reported for a given multipole.}
\label{tab:qnm_wkb_n5}
\end{table*}

Quasinormal modes correspond to solutions satisfying purely ingoing behavior at the horizon and purely outgoing waves at spatial infinity,
\begin{equation}
\Psi \propto
\begin{cases}
e^{-i\omega r_*}, & r_* \to -\infty,\\[4pt]
e^{+i\chi r_*}, & r_* \to +\infty .
\end{cases}
\end{equation}
Following \cite{Konoplya:2004wg}, we choose the sign of $\chi$ so that $\re{\chi}$ and $\re{\omega}$ share the same sign.

For massive fields the structure of the effective potential may differ significantly from the massless case. In particular, when the mass parameter becomes sufficiently large, the potential barrier can lose its local maximum for certain values of the multipole number $\ell$. In such situations the standard WKB approach, which relies on the presence of a single barrier, ceases to be applicable. The quasinormal spectrum must then be determined using alternative techniques such as time-domain integration or other methods applicable to arbitrary potential profiles.

It is important that the effective potential is positive definite everywhere outside the event horizon, both for the massive scalar field considered here, and for gravitational perturbations considered in \cite{Konoplya:2025ect}. Therefore, the differential operator
\begin{equation}
\mathcal{D} = -\frac{\partial^2}{\partial r_*^2} + V(r)
\end{equation}
is a positive self-adjoint operator in the Hilbert space of square integrable functions of \(r_*\). Consequently, solutions of the perturbation equations with compact support initial conditions must be bounded (see, for instance \cite{Abdalla:2006qj,Konoplya:2007jv}), and all quasinormal modes must be decayed.

\section{WKB approach and time-domain integration}
\label{sec:methods}
In this work we employ two methods to determine the quasinormal spectrum: the higher-order WKB approximation supplemented by Padé resummation and time-domain integration. The WKB approach, without Padé approximants, is also used to compute the grey-body factors. Throughout the paper the WKB method serves as the primary computational tool, while the time-domain integration, as much more time consuming, is mainly used to verify the reliability of the WKB results.

\subsection{WKB approach with Pad\'e resummation}

To determine the quasinormal spectrum we employ the higher--order WKB method  \cite{Iyer:1986np,Konoplya:2003ii,Matyjasek:2017psv,Matyjasek:2019eeu}. 
This technique is based on approximating the solution of the wave equation in the vicinity of the peak of the effective potential by matching WKB expansions across the turning points of the potential barrier. 
In practice, the method provides a quantization condition relating the complex frequency $\omega$ to the properties of the potential and its derivatives evaluated at its maximum.

The WKB expansion has been developed to high orders, allowing one to obtain accurate results for black-hole perturbations whenever the effective potential forms a single barrier and the multipole number is not too small. 
The resulting expression for the frequency is written as an asymptotic series whose coefficients depend on successive derivatives of the potential at the extremum. 
While higher orders generally improve the accuracy, the series is not strictly convergent and may exhibit oscillatory behavior when truncated.

To stabilize the approximation we follow the standard procedure of applying Pad\'e resummation to the WKB series. 
Instead of using the truncated expansion directly, the series is transformed into a rational function whose numerator and denominator are polynomials constructed from the WKB coefficients. 
This procedure significantly improves the convergence properties of the approximation and typically yields results that agree well with more precise methods such as the continued-fraction technique.

In the present work we use high-order WKB formulas supplemented by Pad\'e approximants to extract the quasinormal frequencies from the effective potential obtained for the Einasto-supported geometry. 
The method is particularly efficient for the fundamental modes and low overtones, where the potential retains a well-defined barrier shape \cite{Stuchlik:2025ezz,Konoplya:2023moy,Arbelaez:2026eaz,Skvortsova:2024atk,Bolokhov:2023dxq,Kanti:2006ua,Lutfuoglu:2026xlo,Han:2026fpn,Malik:2024tuf,Konoplya:2025hgp,Bolokhov:2024bke,Skvortsova:2023zmj,Konoplya:2006ar,Konoplya:2010vz,Lutfuoglu:2026gis,Konoplya:2002wt,Lambiase:2024lvo,Pantig:2025eda,Gogoi:2024epx,Konoplya:2010kv}. For backgrounds admitting analytic expressions, that is, for $\tilde{n}=1/2$ and $\tilde{n}=1$ we use the WKB expansions up to the 16th orders \cite{Matyjasek:2017psv,Matyjasek:2019eeu}. In cases where the geometry is obtained numerically, the calculation is more time-consuming and we were restricted to the 8th-order WKB approximation. To improve the stability of the series we apply Pad\'e resummation, choosing the approximant such that $\tilde{m}=\tilde{n}$, with $\tilde{m}+\tilde{n}$ equal to the order of the WKB expansion \cite{Matyjasek:2017psv}.

\subsection{Time-domain evolution}

To verify the quasinormal spectrum independently we evolve the perturbations directly in the time domain. 
Introducing the null coordinates
\begin{equation}
u=t-r_*, \qquad v=t+r_* ,
\end{equation}
the wave equation reduces to
\begin{equation}
4\,\frac{\partial^2\Psi}{\partial u\,\partial v}+V(r)\Psi=0 .
\end{equation}

Since the effective potential is naturally expressed as a function of the radial coordinate $r$, while the evolution is performed in terms of the tortoise coordinate $r_*$, the relation
$dr_{*}/dr=f^{-1}(r)$ is used to construct the mapping $r(r_*)$ numerically. The potential entering the evolution equation is therefore evaluated as $V(r(r_*))$ on the computational grid.

The time evolution is performed on a characteristic grid in the $(u,v)$ plane using a finite-difference discretization. 
Denoting the grid points by $S=(u,v)$, $W=(u+\Delta,v)$, $E=(u,v+\Delta)$ and $N=(u+\Delta,v+\Delta)$, the field value at the new point is obtained from
\begin{equation}
\Psi_N=\Psi_W+\Psi_E-\Psi_S-\frac{\Delta^2}{8}
V(S)\left(\Psi_W+\Psi_E\right)+\mathcal{O}(\Delta^4).
\end{equation}

The evolution is initiated by specifying a localized Gaussian wave packet along one of the initial null segments,
\begin{equation}
\Psi(u=u_0,v)=\exp\!\left[-\frac{(r_*-r_{*0})^2}{\sigma^2}\right],
\end{equation}
with the center $r_{*0}$ chosen near the maximum of the effective potential. This choice efficiently excites the dominant quasinormal modes and suppresses the long initial transient stage associated with waves propagating from distant regions.

After a short period of prompt response the waveform typically enters a stage of exponentially damped oscillations dominated by quasinormal modes. The corresponding complex frequencies are extracted from this ringdown signal by fitting the late-time profile. Because this method does not rely on any specific shape of the effective potential, it remains applicable even in situations where the WKB approximation becomes unreliable \cite{Dubinsky:2024gwo,Konoplya:2014lha,Skvortsova:2023zca,Stuchlik:2025mjj,Konoplya:2005et,Skvortsova:2025cah,Cuyubamba:2016cug,Konoplya:2013sba,Dubinsky:2024mwd,Lutfuoglu:2025bsf,Pantig:2022gih,Okyay:2021nnh,Yang:2022ifo,Gogoi:2023fow}.

\begin{figure*}
\resizebox{\linewidth}{!}{\includegraphics{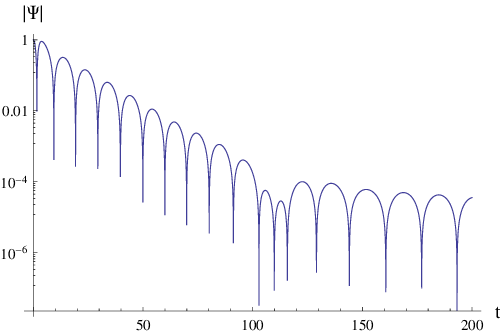}\includegraphics{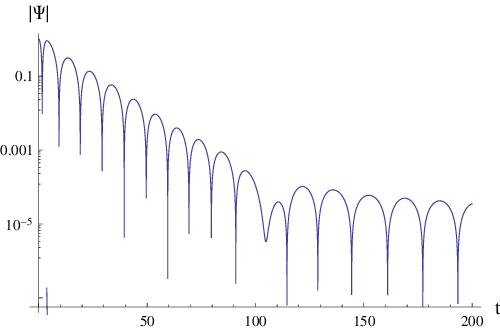}}
\caption{Left: Semi-logarithmic time-domain profile for $\tilde{n}=1/2$, $\ell=1$, $h=1$, and $\mu=0.2$. The ringing period is modified by intermediate asymptotic tails. The QNM obtained by the Prony method from the time-domain integration is $\omega_{n=0} = 0.3084 - 0.0823063 i$. Right: Semi-logarithmic time-domain profile for $\tilde{n}=1$, $\ell=1$, $h=0.38$, and $\mu=0.2$. QNM given by the time-domain profile is $\omega_{n=0} =0.310 - 0.088 i$ on any temporal range inside a wide period from $t=20$, until $t=90$. The difference between the WKB results and time-domain integration may occur owing to the contamination of power-law enveloped tails.}\label{fig:timedomain_nhalf_l1_h1_mu02}
\end{figure*}

\section{Quasinormal modes}\label{sec:QNMs}

Throughout this section, the parameter $\mu$ is interpreted as an effective mass induced by an external magnetic field for an otherwise massless scalar mode near the black hole. Following the standard magnetized-background treatment \cite{Konoplya:2007yy,Konoplya:2008hj,Wu:2015fwa,Davlataliev:2024mjl,Kokkotas:2010zd}, we use
\begin{equation}
\mu^2 = 4 B^2 m^2,
\end{equation}
where $m$ is the azimuthal quantum number, which appears during separation of variables, and $B$ is the strength of the external magnetic field.   
Therefore, for each multipole sector, the magnetic field shifts the effective asymptotic threshold of the potential and changes both the leakage rate through the barrier and the oscillation rate, which is why it affects QNMs and grey-body factors simultaneously.

The approximation employed in our work corresponds to the weak-field regime $B^2 M^2 \ll 1$, in which the back-reaction of the magnetic field on the Schwarzschild lapse function $f(r)$ can be neglected and only the leading $B^2$ corrections to the effective potential are retained. 
In \cite{Kokkotas:2010zd} it was shown that the effect of the magnetic field upon quasinormal spectrum can be obtained via discarding the non-asymptotically flat terms, that is, by neglecting terms of order higher than $B^2$ in the wave equation.
The approximation is expected to break down once $B^2 M^2 \sim 1$, when higher-order corrections to both the metric and the wave equation become non-negligible. However, the energy density of the magnetic field surrounding the black hole is typically much smaller than the characteristic gravitational energy density and decreases rapidly with distance. Therefore, the assumption that the magnetic field can be modeled as an effective mass is well justified.

The numerical data in Tables~\ref{tab:qnm_wkb_comparison_nhalf} and \ref{tab:qnm_wkb_comparison} show that the Einasto parameters strongly affect both the oscillation frequency and the damping rate. For both regular solutions, $\tilde n=1/2$ and $\tilde n=1$, increasing the halo scale parameter $h$ generally increases $\re{\omega}$ and decreases $|\im{\omega}|$, i.e. the oscillations become faster and less damped. This tendency is most stable for $\ell=1,2$, where the WKB approximants remain close over a broad range of $h$. In terms of the effective potential, this behavior is consistent with Figs.~\ref{fig:potentials_all_nhalf} and \ref{fig:potentials_all}: changes in the barrier shape modify both the oscillation timescale (through $\re{\omega}$) and the leakage rate (through $\im{\omega}$).

The time-domain profile in Fig.~\ref{fig:timedomain_nhalf_l1_h1_mu02} provides an additional consistency check for the frequency-domain results. From that profile we extract $\omega_{\rm TD}\approx 0.3084-0.08231i$, while the corresponding WKB values in Table~\ref{tab:qnm_wkb_comparison_nhalf} for $\tilde n=1/2$, $\ell=1$, and nearby $\mu=0.2$ are $\omega_{\rm WKB16}=0.307463-0.080345i$ and $\omega_{\rm WKB14}=0.307493-0.080011i$. Thus, the real part agrees at the sub-percent level, whereas the damping rate differs by only a few percent, which is acceptable for this level of approximation and confirms the same physical trend toward weaker damping with increasing effective mass.

Notice that the asymptotic tails contaminate the signal at progressively earlier times as the field mass $\mu$ increases. Unfortunately, there is no universal remedy for this problem. In practice, one can reliably extract quasinormal frequencies only in the regime where the multipole number $\ell$ is not too small and the field mass $\mu$ is not too large. In this case, the ringdown stage remains sufficiently long to allow for an accurate frequency extraction.

A key physical trend is the approach of the damping rate to zero, indicating the onset of quasi-resonances. This is clearly visible in Table~\ref{tab:qnm_wkb_n5}. For $\ell=2$ at $\tilde n=5$, one has $|\im{\omega}|\approx 9.14\times10^{-2}$ at $\mu=0$, while at $\mu=1.0$ it drops to $|\im{\omega}|\approx 8.69\times10^{-3}$, i.e. by about one order of magnitude, whereas $\re{\omega}$ increases from $0.706753$ to $0.877380$. For $\ell=1$, in the available range up to $\mu=0.6$, the damping also decreases substantially, from $\approx 9.18\times10^{-2}$ to $\approx 3.94\times10^{-2}$. Therefore, the data show a clear tendency toward long-lived modes for sufficiently large effective mass.

The difference between WKB orders is not always tiny, but in most cases it remains smaller than the total spectral variation caused by nonzero $\mu$ and by background deformation. From Tables~\ref{tab:qnm_wkb_comparison_nhalf} and \ref{tab:qnm_wkb_comparison}, the relative WKB-order difference is usually at the sub-percent level (often much smaller), with only a few less favorable points showing larger deviations. At the same time, the mass-driven and geometry-driven shifts in frequency and damping are much stronger, especially near the quasi-resonant regime where $|\im{\omega}|$ is strongly suppressed. This indicates that the observed trend is physical and not merely a truncation artifact of the WKB expansion.

Summarizing the discussion of accuracy, we may regard as reliable only those values for which the relative difference between the 14th- and 16th-order results remains within one percent, while differences of several percent or more indicate instability of the WKB method. However, a recent study \cite{Konoplya:2026rjh} indicates that applying the WKB method at very high orders, reaching several hundreds, may remedy this problem and allow one to approach the quasi-resonant regime.

Overall, the trend toward weak damping should be interpreted with an important qualification. Strictly speaking, quasi-resonances (arbitrarily small $|\im{\omega}|$ at finite $\re{\omega}$) are approached when the effective mass is close to the critical value for which the radial potential no longer has a barrier maximum. In that regime, the usual barrier-leakage picture changes qualitatively, and the mode lifetime can become very large.

This also explains why the QNM tables are truncated at different maximal $\mu$ for different multipoles. The critical mass depends on $\ell$: larger $\ell$ increases the centrifugal contribution and keeps the potential maximum for higher values of $\mu$. Consequently, calculations for higher multipoles can be extended to larger effective masses before the peak disappears, whereas lower-$\ell$ modes reach the critical regime earlier. The observed suppression of damping in our data is therefore consistent with approaching this $\ell$-dependent critical point.

\begin{figure*}
\resizebox{\linewidth}{!}{\includegraphics{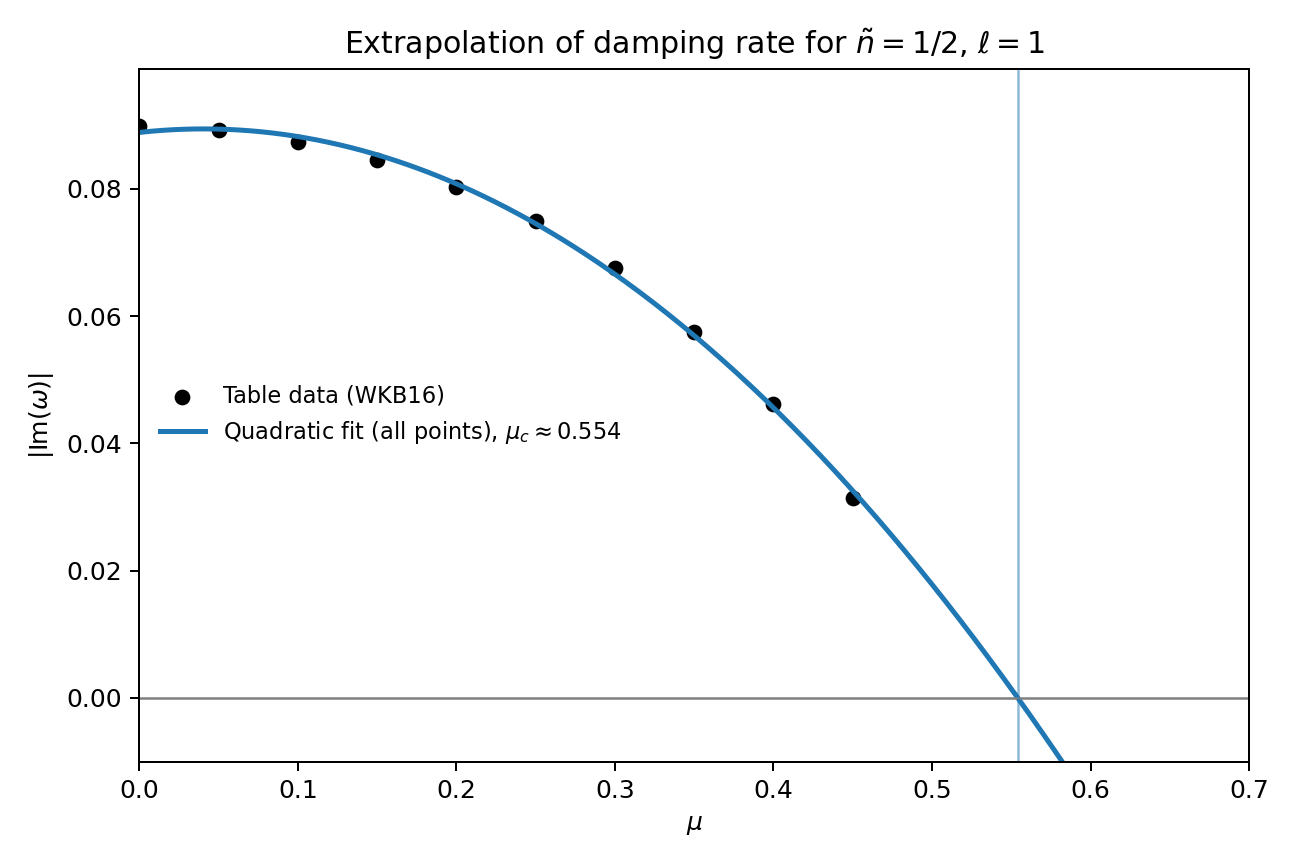}\includegraphics{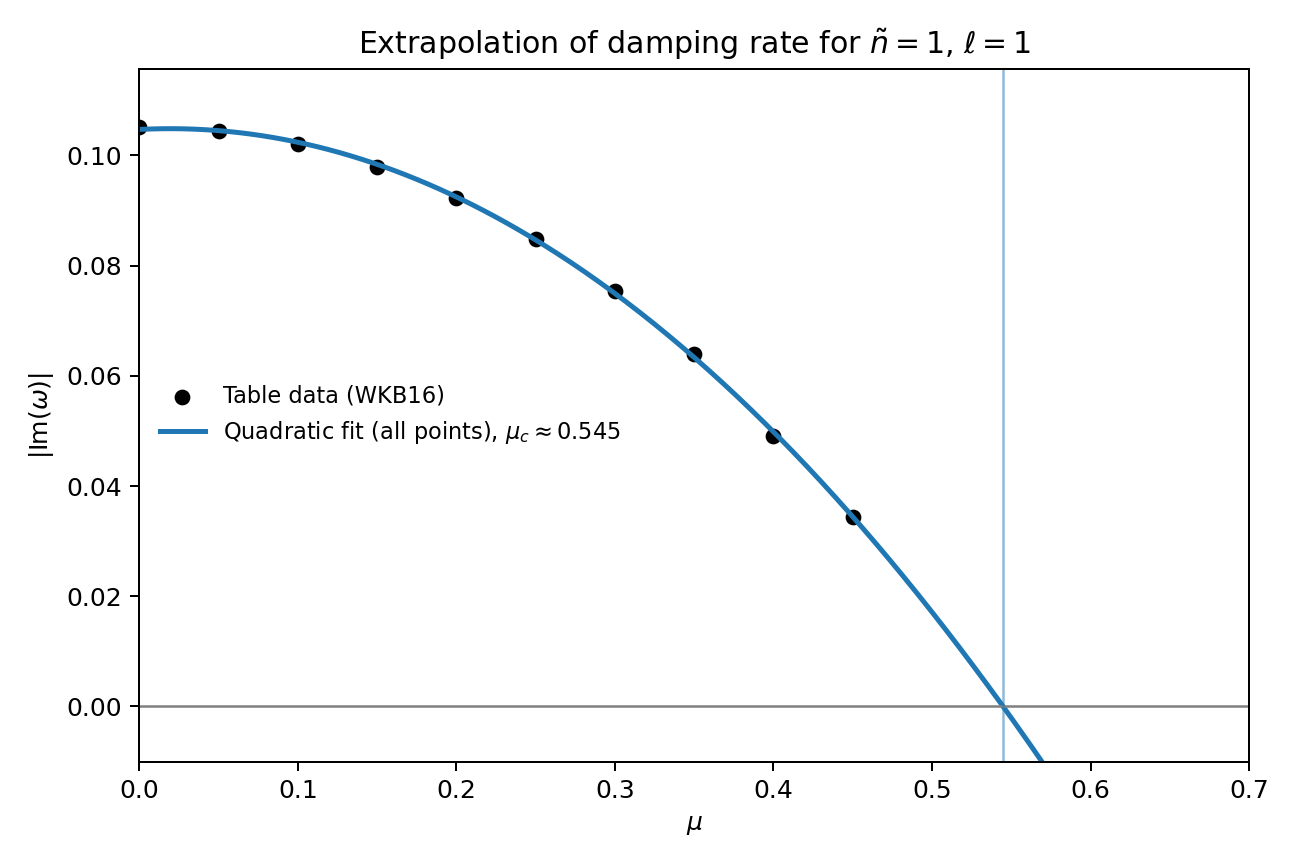}}
\caption{Extrapolation of the damping rate $|\im{\omega}|$ versus effective mass $\mu$ for the fundamental scalar mode with $\ell=1$, using only tabulated WKB16 data. Left: $\tilde n=1/2$ (Table~\ref{tab:qnm_wkb_comparison_nhalf}), yielding $\mu_c\approx0.554$. Right: $\tilde n=1$ (Table~\ref{tab:qnm_wkb_comparison}), yielding $\mu_c\approx0.545$. The solid curves are quadratic fits to the tabulated points.}\label{fig:qnm_imag_extrapolation_models}
\end{figure*}

To illustrate this tendency directly from tabulated values, we fitted the WKB16 data for $|\im{\omega}|$ at $\ell=1$ with a quadratic function of $\mu$ for the $\tilde n=1/2$ and $\tilde n=1$ models. In both cases the extrapolated damping rate reaches zero at finite $\mu$, indicating an approach to the critical regime where the barrier peak disappears. This behavior is shown in Fig.~\ref{fig:qnm_imag_extrapolation_models}, where the fitted curves cross $|\im{\omega}|=0$ at $\mu_c\approx0.554$ for $\tilde n=1/2$ and $\mu_c\approx0.545$ for $\tilde n=1$. Although these values are extrapolated, they provide a quantitative estimate of where long-lived modes are expected to emerge for $\ell=1$.

\section{Grey-body factors}
\label{sec:gbf}

The grey-body factor characterizes the probability that radiation produced near the horizon penetrates the effective potential barrier and reaches spatial infinity. For black-hole perturbations whose effective potential forms a single barrier, the transmission coefficient can be evaluated efficiently using the WKB approximation.

Within this framework the reflection and transmission coefficients take the form
\begin{equation}
1-|R|^2=\frac{\Omega}{\chi}|T|^2=\left(1+e^{2\pi i K}\right)^{-1},
\end{equation}
where
\begin{equation}
iK=\frac{\Omega^2-V_0}{\sqrt{-2V_0''}}+\sum_{i=2}^{N}\Lambda_i .
\end{equation}
Here $V_0$ denotes the height of the effective potential, $V_0''$ is its second derivative with respect to the tortoise coordinate at the maximum, and $\Lambda_i$ represent higher-order WKB corrections determined by successive derivatives of the potential. This approach was frequently used for finding sufficiently accurate values of grey-body factors \cite{Konoplya:2023ahd,Konoplya:2009hv,Lutfuoglu:2025eik,Dubinsky:2025wns,Dubinsky:2025ypj,Malik:2025qnr,Malik:2025erb}.

A useful relation emerges between the transmission coefficient and the quasinormal spectrum of the system. In particular, the grey-body factor for multipole number $\ell$ can be expressed through the lowest quasinormal frequencies,
\cite{Konoplya:2024vuj,Konoplya:2024lir}
\begin{equation}
\Gamma_\ell(\Omega)=\frac{1}{1+e^{2\pi iK}},
\end{equation}
where the quantity $K$ may be expanded in terms of the real frequency $\Omega$ together with the fundamental quasinormal mode $\omega_0$ and the first overtone $\omega_1$,
\begin{eqnarray}\nonumber
&&i  K=\frac{\Omega^2-\re{\omega_0}^2}{4\re{\omega_0}\im{\omega_0}}\Biggl(1+\frac{(\re{\omega_0}-\re{\omega_1})^2}{32\im{\omega_0}^2}
\\\nonumber&&\qquad\qquad-\frac{3\im{\omega_0}-\im{\omega_1}}{24\im{\omega_0}}\Biggr)
-\frac{\re{\omega_0}-\re{\omega_1}}{16\im{\omega_0}}
\\\nonumber&& -\frac{(\Omega^2-\re{\omega_0}^2)^2}{16\re{\omega_0}^3\im{\omega_0}}\left(1+\frac{\re{\omega_0}(\re{\omega_0}-\re{\omega_1})}{4\im{\omega_0}^2}\right)
\\\nonumber&& +\frac{(\Omega^2-\re{\omega_0}^2)^3}{32\re{\omega_0}^5\im{\omega_0}}\Biggl(1+\frac{\re{\omega_0}(\re{\omega_0}-\re{\omega_1})}{4\im{\omega_0}^2}
\\\nonumber&&\qquad +\re{\omega_0}^2\Biggl(\frac{(\re{\omega_0}-\re{\omega_1})^2}{16\im{\omega_0}^4}
\\&&\qquad\qquad -\frac{3\im{\omega_0}-\im{\omega_1}}{12\im{\omega_0}}\Biggr)\Biggr)+ \Order{\frac{1}{\ell^3}}.
\label{eq:gbsecondorder}
\end{eqnarray}

In the eikonal regime ($\ell\gg1$) this correspondence becomes exact for a broad class of single-barrier potentials. For moderate values of $\ell$ the relation remains approximate but typically provides accurate estimates for the transmission probability. Numerous recent analyses confirm that this connection reproduces grey-body factors with high precision across a wide range of black-hole spacetimes and field perturbations
\cite{Lutfuoglu:2025blw,Malik:2024cgb,Han:2025cal,
Skvortsova:2024msa,Malik:2025dxn,Lutfuoglu:2025hjy,Dubinsky:2025nxv,
Bolokhov:2024otn,Lutfuoglu:2025ldc,Lutfuoglu:2025ohb}.

However, the applicability of the WKB-based correspondence is limited to cases where the effective potential possesses a well-defined barrier. In certain modified gravity theories, particularly those involving higher-curvature corrections, the eikonal form of the potential may deviate significantly from the standard centrifugal structure. In such situations the WKB expansion becomes unreliable and the analytic relation between grey-body factors and quasinormal modes may fail
\cite{Konoplya:2017ymp,Konoplya:2017zwo,Konoplya:2017lhs}. Similar difficulties may arise when the matching between the near-peak WKB expansion and the asymptotic regions becomes inaccurate
\cite{Bolokhov:2023dxq}.

\begin{figure*}
\resizebox{\linewidth}{!}{\includegraphics{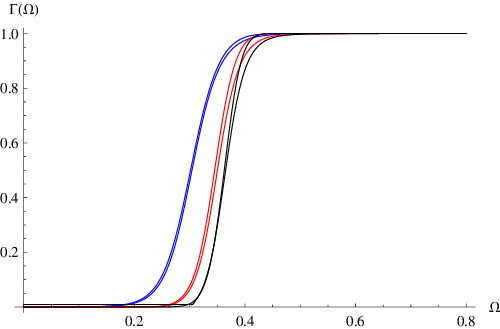}\includegraphics{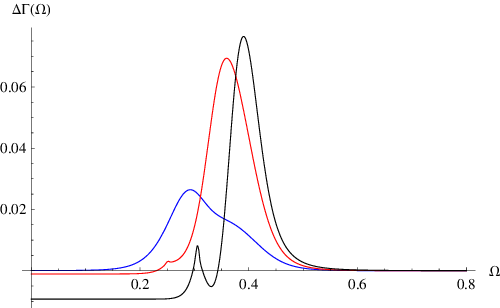}}
\caption{Grey-body factors for $\ell=1$, $\tilde{n}=1$, $M=1$, and $\mu=0$ (blue), $0.3$ (red), and $0.35$ (black) computed using the WKB method and via the correspondence with quasinormal modes (QNMs), together with their difference.}\label{fig:GBFL1}
\end{figure*}

\begin{figure*}
\resizebox{\linewidth}{!}{\includegraphics{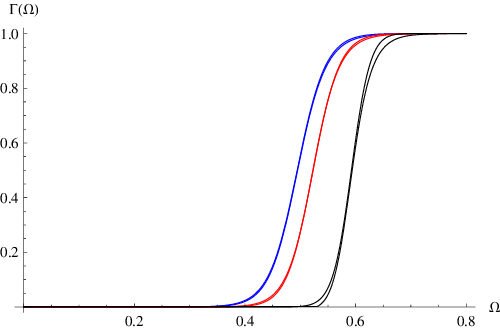}\includegraphics{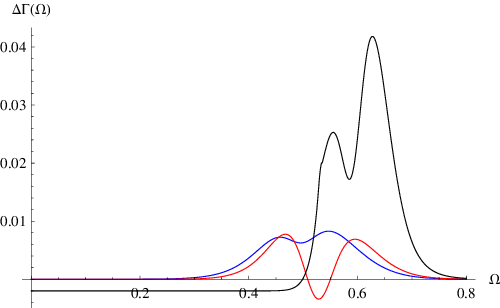}}
\caption{Grey-body factors for $\ell=2$, $\tilde{n}=1$, $M=1$, and $\mu=0$ (blue), $0.3$ (red), and $0.55$ (black) computed using the WKB method and via the correspondence with quasinormal modes (QNMs), together with their difference.}\label{fig:GBFL2}
\end{figure*}

The numerical results shown in Figs.~\ref{fig:GBFL1} and \ref{fig:GBFL2} confirm that, for the considered parameter ranges, the WKB transmission coefficients and the QNM-based reconstruction are in close agreement. The lower panels (difference plots) remain comparatively small across most of the frequency interval, indicating that the correspondence captures the main frequency dependence of the transmission probability not only in the strict eikonal limit but also for the moderate multipoles used here.

Notice that the correspondence between quasinormal modes and grey-body factors relies crucially on the applicability of the WKB approximation to both problems. Therefore, for sufficiently large field masses, when the effective potential no longer possesses a maximum (see figs. \ref{fig:potentials_all_nhalf} and \ref{fig:potentials_all}), the WKB approach breaks down and, consequently, the correspondence ceases to hold, in agreement with similar observations reported recently in \cite{Lutfuoglu:2025kqp,Bolokhov:2026uol,Malik:2026jzl} for other configurations.

When the effective potential develops a double-peak structure \cite{Guo:2022ghl,Konoplya:2025bte}, the correspondence may cease to be valid or remain only approximate with potentially large errors, particularly when the secondary peak is relatively small. However, as can be seen from the effective-potential profiles for the range of parameters considered here, no such double-peak structure arises.

A clear physical trend is also visible when the effective mass $\mu$ is increased: the transmission is suppressed at low and intermediate frequencies and the onset of efficient transmission shifts toward larger $\Omega$. This behavior is consistent with the effective potential picture discussed above: increasing $\mu$ raises the asymptotic part of the potential and reduces penetration through the barrier at fixed frequency. At the same time, for sufficiently large $\Omega$, all curves approach the geometric-optics regime with high transmission, as expected for single-barrier scattering.

\begin{figure}
\resizebox{\linewidth}{!}{\includegraphics{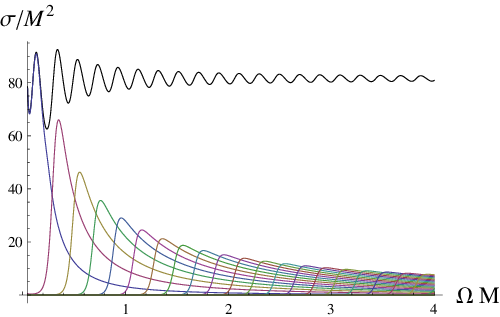}}
\caption{Partial and total absorption cross-sections for the scalar field in the Einasto-supported regular black-hole model with $\tilde{n}=1$ and $\mu=0.1$ ($M=1$). }\label{fig:absorption_n1_mu01}
\end{figure}

From the grey-body factors one obtains the absorption cross-section. For a minimally coupled scalar field in four dimensions, the partial absorption cross-section for multipole $\ell$ is
\begin{equation}
\sigma_{\ell}(\Omega)=\frac{\pi}{\Omega^2}(2\ell+1)\,\Gamma_{\ell}(\Omega),
\end{equation}
and the total absorption cross-section is the sum over multipoles,
\begin{equation}
\sigma_{\mathrm{abs}}(\Omega)=\sum_{\ell=0}^{\infty}\sigma_{\ell}(\Omega)
=\frac{\pi}{\Omega^2}\sum_{\ell=0}^{\infty}(2\ell+1)\,\Gamma_{\ell}(\Omega).
\end{equation}
The resulting partial and total cross-sections for $\tilde n=1$ and $\mu=0.1$ are presented in Fig.~\ref{fig:absorption_n1_mu01}. The total cross-section reproduces the expected smooth transition from the low-frequency suppressed regime to the higher-frequency regime where more partial waves contribute efficiently.

\vspace{5mm}
\section{Conclusion}\label{sec:conc}

In this work we analyzed massive scalar perturbations in regular black-hole geometries generated by Einasto matter distributions, with emphasis on three observables: quasinormal modes, grey-body factors, and absorption cross-sections. The combined WKB--Pad\'e and time-domain analysis shows that the Einasto parameters and the effective mass $\mu$ produce substantial and systematic changes in the spectrum.

For the quasinormal frequencies, the dominant trend is a decrease of $|\im{\omega}|$ as the effective mass increases, accompanied by a shift of $\re{\omega}$. In the $\tilde n=5$ case this leads to a clear quasi-resonant regime, where damping becomes strongly suppressed and modes become long-lived. Although the discrepancy between different WKB orders is not uniformly tiny across the whole parameter space, it is generally smaller than the total physical shift induced by nonzero $\mu$, supporting the robustness of the observed spectral trends.

For scattering, the grey-body factors computed by direct WKB transmission and by the QNM-based correspondence are in good agreement for the considered multipoles and parameter ranges. The transmission curves show the expected behavior under increasing $\mu$: stronger low-frequency suppression and a shift of the efficient-transmission region toward higher frequencies. From these coefficients, the partial and total absorption cross-sections display the corresponding transition from a suppressed low-frequency regime to enhanced high-frequency absorption.

Overall, our results indicate that regular-core structure and environmental effects encoded by the Einasto profile leave a consistent and potentially observable imprint on both ringdown and wave-scattering characteristics. This supports the use of quasinormal and scattering data as complementary probes of non-vacuum regular black-hole geometries. Because $\mu^2\propto B^2 m^2$, these signatures can also be interpreted as indirect probes of magnetic-field strength near the black hole: stronger magnetic fields push the system toward longer-lived ringdown modes and delayed transmission onset in frequency-domain scattering.

\vspace{5mm}
\textbf{Data availability statement:} No new data was generated during this research.

\vspace{5mm}
\acknowledgments
I would like to thank Kirill Bronnikov for useful discussions.

\bibliography{BHEinasto}
\end{document}